\documentclass[12pt]{article}

\usepackage{mathrsfs}
\usepackage[T1]{fontenc}
\usepackage{mathpazo}
\usepackage{setspace}
\usepackage{amsfonts}
\usepackage{amssymb}
\usepackage{amsmath}
\usepackage{epsfig}
\usepackage{latexsym}
\usepackage{color}
\usepackage{graphicx}
\usepackage{nicefrac}
\usepackage[latin1]{inputenc}
\usepackage{pstricks}
\usepackage{slashed}
\usepackage{multirow}

\usepackage{hyperref}

\def\hybrid{\topmargin -30pt    \oddsidemargin 0pt 
        \headheight 0pt \headsep 0pt
        \textwidth 6.25in       
        \textheight 9.5in       
        \marginparwidth .875in
        \parskip 5pt plus 1pt   \jot = 1.5ex}

\hybrid

\def\baselinestretch{1.2}

\catcode`\@=11

\def\marginnote#1{}
\newcount\hour
\newcount\minute
\newtoks\amorpm
\hour=\time\divide\hour by60
\minute=\time{\multiply\hour by60 \global\advance\minute by-\hour}
\edef\standardtime{{\ifnum\hour<12 \global\amorpm={am}%
        \else\global\amorpm={pm}\advance\hour by-12 \fi
        \ifnum\hour=0 \hour=12 \fi
        \number\hour:\ifnum\minute<10 0\fi\number\minute\the\amorpm}}
\edef\militarytime{\number\hour:\ifnum\minute<10 0\fi\number\minute}

\def\draftlabel#1{{\@bsphack\if@filesw {\let\thepage\relax
   \xdef\@gtempa{\write\@auxout{\string
      \newlabel{#1}{{\@currentlabel}{\thepage}}}}}\@gtempa
   \if@nobreak \ifvmode\nobreak\fi\fi\fi\@esphack}
        \gdef\@eqnlabel{#1}}
\def\@eqnlabel{}
\def\@vacuum{}
\def\draftmarginnote#1{\marginpar{\raggedright\scriptsize\tt#1}}

\def\draft{\oddsidemargin -.5truein
        \def\@oddfoot{\sl preliminary draft \hfil
        \rm\thepage\hfil\sl\today\quad\militarytime}
        \let\@evenfoot\@oddfoot \overfullrule 3pt
        \let\label=\draftlabel
        \let\marginnote=\draftmarginnote
   \def\@eqnnum{(\theequation)\rlap{\kern\marginparsep\tt\@eqnlabel}%
\global\let\@eqnlabel\@vacuum}  }

\def\draft2{
        \def\@oddfoot{\sl preliminary draft \hfil
        \rm\thepage\hfil\sl\today\quad\militarytime}
        \let\@evenfoot\@oddfoot \overfullrule 3pt
        \let\label=\draftlabel
        \let\marginnote=\draftmarginnote
   \def\@eqnnum{(\theequation)\rlap{\kern\marginparsep\tt\@eqnlabel}%
\global\let\@eqnlabel\@vacuum}  }

\def\preprint{\twocolumn\sloppy\flushbottom\parindent 2em
        \leftmargini 2em\leftmarginv .5em\leftmarginvi .5em
        \oddsidemargin -.5in    \evensidemargin -.5in
        \columnsep .4in \footheight 0pt
        \textwidth 10.in        \topmargin  -.4in
        \headheight 12pt \topskip .4in
        \textheight 6.9in \footskip 0pt
        \def\@oddhead{\thepage\hfil\addtocounter{page}{1}\thepage}
        \let\@evenhead\@oddhead \def\@oddfoot{} \def\@evenfoot{} }

\def\numberbysection{\@addtoreset{equation}{section}
        \def\theequation{\thesection.\arabic{equation}}}

\def\underline#1{\relax\ifmmode\@@underline#1\else
        $\@@underline{\hbox{#1}}$\relax\fi}

\def\titlepage{\@restonecolfalse\if@twocolumn\@restonecoltrue\onecolumn
     \else \newpage \fi \thispagestyle{empty}\c@page\z@
        \def\thefootnote{\fnsymbol{footnote}} }

\def\endtitlepage{\if@restonecol\twocolumn \else \newpage \fi
        \def\thefootnote{\arabic{footnote}}
        \setcounter{footnote}{0}}  

\catcode`@=12
\relax

\def\figcap{\section*{Figure Captions\markboth
        {FIGURECAPTIONS}{FIGURECAPTIONS}}\list
        {Figure \arabic{enumi}:\hfill}{\settowidth\labelwidth{Figure
999:}
        \leftmargin\labelwidth
        \advance\leftmargin\labelsep\usecounter{enumi}}}
 \relax
\def\tablecap{\section*{Table Captions\markboth
        {TABLECAPTIONS}{TABLECAPTIONS}}\list
        {Table \arabic{enumi}:\hfill}{\settowidth\labelwidth{Table
999:}
        \leftmargin\labelwidth
        \advance\leftmargin\labelsep\usecounter{enumi}}}
 \relax
\def\reflist{\section*{References\markboth
        {REFLIST}{REFLIST}}\list
        {[\arabic{enumi}]\hfill}{\settowidth\labelwidth{[999]}
        \leftmargin\labelwidth
        \advance\leftmargin\labelsep\usecounter{enumi}}}
 \relax
\makeatletter
\newcounter{pubctr}
\def\publist{\@ifnextchar[{\@publist}{\@@publist}}
\def\@publist[#1]{\list
        {[\arabic{pubctr}]\hfill}{\settowidth\labelwidth{[999]}
        \leftmargin\labelwidth
        \advance\leftmargin\labelsep
        \@nmbrlisttrue\def\@listctr{pubctr}
        \setcounter{pubctr}{#1}\addtocounter{pubctr}{-1}}}
\def\@@publist{\list
        {[\arabic{pubctr}]\hfill}{\settowidth\labelwidth{[999]}
        \leftmargin\labelwidth
        \advance\leftmargin\labelsep
        \@nmbrlisttrue\def\@listctr{pubctr}}}
 \relax
\makeatother

\def\be{\begin{equation}}
\def\ee{\end{equation}}
\def\ba{\begin{eqnarray}}
\def\ea{\end{eqnarray}}

\def\del{\partial}

\def\k{\kappa}
\def\r{\rho}
\def\a{\alpha}

\def\b{\beta}

\def\g{\gamma}
\def\G{\Gamma}
\def\d{\delta}
\def\D{\Delta}
\def\e{\epsilon}

\def\m{\mu}
\def\n{\nu}

\def\Om{\Omega}
\def\l{\lambda}
\def\L{\Lambda}
\def\s{\sigma}

\def\cP{{\cal P}}

\def\no{\noindent}

\def\qq{\qquad}

\def\IR{\relax{\rm I\kern-.18em R}}

\def\inv{^{\raise.0ex\hbox{${\scriptscriptstyle -}$}\kern-.05em 1}}

\def \ha {{\frac{1}{2}}}

\def \ov {\over}

\def\diag{{\rm diag}}

\begin{document}


\renewcommand{\theequation}{\thesection.\arabic{equation}}
\csname @addtoreset\endcsname{equation}{section}

\begin{titlepage}
\begin{center}

\hfill KIAS P11015  \\
\hfill FPAUO-11/04  \\

\phantom{xx}
\vskip 0.5in

{\large \bf  Non-abelian T-duality, Ramond Fields and Coset Geometries}

\vskip 0.5in

{\bf Yolanda  Lozano}${}^{1a}$,\phantom{x} {\bf Eoin  \'O Colg\'ain}${}^{2b}$,
\\
\vskip .15 cm

{\bf Konstadinos Sfetsos}${}^{3c}$\phantom{x}  and \phantom{x} {\bf Daniel C.
Thompson}${}^{4d}$ \vskip 0.1in

\vskip .2in

${}^1$Department of Physics,  University of Oviedo,\\
Avda.~Calvo Sotelo 18, 33007 Oviedo, Spain

\vskip .2in

${}^2$Korea Institute for Advanced Study,  \\
Dongdaemun-gu, Seoul 130-722, Korea \\

\vskip .2in

${}^3$Department of Engineering Sciences, University of Patras,\\
26110 Patras, Greece\\

\vskip .2in

${}^4$Theoretische Natuurkunde, Vrije Universiteit Brussel, and \\
The International Solvay Institutes\\
Pleinlaan 2, B-1050, Brussels, Belgium \\

\end{center}

\vskip .4in

\centerline{\bf Abstract}

\no
We extend previous work on non-abelian T-duality in the presence of
Ramond fluxes to cases in which the duality group acts with isotropy such as in backgrounds containing
coset spaces. In the process we generate new supergravity solutions related to D-brane configurations and to
standard supergravity compactifications.

\vfill
\no
 {
$^a$ylozano@uniovi.es,
 $^b$eoin@kias.re.kr, $^c$sfetsos@upatras.gr,
$^d$dthompson@tena4.vub.ac.be.}

\end{titlepage}
\vfill
\eject


\tableofcontents

\def\baselinestretch{1.2}
\baselineskip 20 pt

\section{Introduction}

The idea of extending abelian T-duality \cite{Buscher:1987sk} to non-abelian isometry groups has a long
history 
\cite{delaossa:1992vc}-\cite{Sfetsos:1996pm}. The only true similarity between the two cases is the fact
that both can be given a path integral formulation. However, there are a number of notable differences
that clearly distinguish the two cases.
Unlike the abelian case, when the isometries are non-commuting, they
are no longer present in the T-dual background and the
transformation is non-invertible in a path integral approach.
Additionally, in general, one cannot establish non-abelian
duality as an exact equivalence between partition functions.
 Nonetheless, such a transformation can
 still have powerful applications as a solution generating technique in supergravity.
Also, in the examples that have been constructed, even if the original
non-abelian group $G$ is compact, the associate variables of the
T-dual background are non-compact. The last remark, together with
some earlier observation in \cite{Sfetsos:1994vz} and technical
advancements in dealing with backgrounds lacking manifest isometries
\cite{Polychronakos:2010fg}, led recently to an improvement of our
understanding. In particular, it was realized that non-abelian
T-duality in pure NS backgrounds can be thought of as describing
infinitely large spin sectors of a parent theory \cite{Polychronakos:2010hd}.
When in the latter's theory $\s$-model the target
space coordinates undergo a stretching or
contraction one obtains the T-dual $\s$-model we are interested in.

\no
In some sense, the situation is similar to fermionic T-duality
\cite{Berkovits:2008ic} which provided an explanation of the dual
superconformal symmetry of ${\cal N}=4$ SYM when applied to $AdS_5 \times S^5$,
which also is not an exact symmetry.
This development motivates a reconsideration of non-abelian T-duality,
in the context of geometries supported by Ramond (RR) fluxes.
In \cite{Sfetsos:2010uq}, non-abelian T-duality was considered for target spaces
 which included some group manifold, $G$, as a subspace and whose curvature was
 supported by RR fluxes. These theories possess a $G_L \times G_R$ isometry group
 and it was shown how to implement the non-abelian
 duality with respect to the $G_L$ symmetry. These situations can naturally
 occur in the near horizon geometries of D-brane configurations.
An example
is the case of $AdS_3 \times S^3 \times T^4$; here a dualisation with
respect to an $SU(2)_L$ symmetry of the $S^3$ results in a solution of massive IIA supergravity.
 Performing a similar dualisation on an $SU(2) \subset SO(6)$ isometry
 for the case of $AdS_5 \times S^5$   gave rise to a solution whose M-theory
 lift captures generic features of the geometries proposed in \cite{Gaiotto:2009gz} (for similar
 geometries constructed in type-IIA see \cite{ReidEdwards:2010qs})
 as gravity duals to  ${\cal N}= 2$ gauge theories.

\no
The formulation of non-abelian T-duality in the presence of Ramond fluxes in \cite{Sfetsos:2010uq}
overcame certain technical difficulties.
To appreciate it, recall that in the abelian case the unique dimensional reduction to nine dimensions
of the type-II supergravities
provided for the transformation rules \cite{Bergshoeff:1995as}.
However, in non-abelian cases an approach along these lines seems more demanding and hasn't been explored so far.
Following this work, it is natural to ask whether the situation can be generalized further to include the case
where the isometry is realized via a coset manifold. For instance,
one may consider, as we indeed do in a particular example,
the dualization of the entire $SO(6)$ isometry that acts on the five-sphere
within $AdS_5 \times S^5$. This is a rather non-trivial extension at both the
technical and conceptual levels.

\no
The aim of this paper is to address exactly this situation and to provide a
whole class of new examples of non-abelian T-dual backgrounds by considering
 target spaces containing coset manifolds.  More precisely, for target spaces
  containing a coset $G/H$ manifold we will perform a duality with respect to
  the full $G$ isometry group and demonstrate how the Ramond fluxes transform
  under the duality.  We illustrate this by providing several examples of
  dualisation in interesting supergravity backgrounds detailed in Table 1.
\begin{table}[h!]
\centering
\begin{tabular}{|c|c|c|}
\hline
Background & Coset & Group Dualised\\
\hline\hline
$AdS_3\times {\bf S^3} \times T^4$  & $SO(4)/SO(3)$ & $SO(4)$ \\
$AdS_5\times {\bf S^5}  $  &$SO(6)/SO(5)$& $SO(6)$ \\
$AdS_4\times {\bf CP^2} \times S^2 $ & $SU(3)/(SU(2) \times U(1)) $& $SU(3) $ \\
$AdS_4\times {\bf S^2 \times S^2 \times S^2 \bf} $ & $(SU(2)/SO(2))^3 $& $SU(2)^3 $ \\
$AdS_4\times {\bf CP^3} $ & $SU(4)/ (SU(3) \times U(1))$ & $SU(4) $ \\
\hline
\end{tabular}
\caption{{\it Examples studied; the relevant coset manifold shown in bold.}}
\label{Table 1}
\end{table}

\no
Unlike the case of group manifolds, the $G$ isometry group typically acts
on the coset $G/H$ with isotropy and it is this feature that introduces
 some technical challenges. This is very evident in the Buscher procedure
 in which the $\dim(G)$ isometry group is gauged; one will have $\dim(G)$
  Lagrange multipliers enforcing a flat connection. Among all these variables
$\dim(G/H)$ will become the T-dual coordinates and the remainder will be gauge fixed.
We will exploit the fact that the dual geometry can
   be parametrised by $H$ invariant combinations of the Lagrange
   multipliers to address this issue and to provide simplified geometries
   produced by dualisation.  Expanding the techniques of \cite{Sfetsos:2010uq}
   we are able to construct the full Ramond fluxes required to support these
   geometries as supergravity solutions which we summarise in table 2.
    A general feature is that the chirality of the dual theory changes
    when $\dim(G)$ is odd and is preserved when this is
     even. One may also see that in all of the dual backgrounds
     there is no $NS$ two-form, something attributable to the fact that the coset spaces are symmetric
and the group we dualized with is the maximal symmetry group (of the
corresponding factor in bold in table 2).

\begin{table}[h!t]
\centering
\begin{tabular}{|c|c|c|}
\hline
Initial Background & Initial RR-Fields & T-Dual RR-Fields\\
\hline\hline
$AdS_3\times {\bf S^3} \times T^4$  & $F_3$ & $F_1\ , \ F_5$ \\
$AdS_5\times {\bf S^5}  $  & $F_5$   &  $F_2$  \\
$AdS_4\times {\bf CP^2} \times S^2 $ & $F_2\ , \ F_4  $& $  F_2\ , \ F_4$  \\
$AdS_4\times {\bf S^2 \times S^2 \times S^2 \bf} $ & $F_2\ , \ F_4  $ & $F_3$  \\
$AdS_4\times {\bf CP^3} $ & $F_2\ , \ F_4 $ & $ F_3 $ \\
\hline
\end{tabular}
\caption{{\it Initial and T-dual backgrounds with the corresponding Ramond fluxes indicated.}}
\label{Table 2}
\end{table}

\no
The structure of the rest of this paper is as follows:
In section 2 we review the general strategy of T-duality in
 the presence of Ramond fields and then in section 3 we show how
  this may be applied to the coset geometries in general. In section 4
  we then present the explicit examples studied.  Due to its additional
  complexity we leave the case of $AdS_4\times CP^3$ as an appendix A to the main article.
We have also included appendix B with useful information on the geometry and Killing vectors
of group and coset spaces,
appendix C with the action of the spinor-Lorentz-Lie derivative on the Killing vectors of the
$AdS_5\times S^5$ space and appendix D on the Killing vectors of $S^5$ as a coset space and the proof
of a useful identity.

\section{General strategy }

Given a supergravity background, in order to perform the non-abelian T-duality transformation we first
allocate the group of isometries with respect to which we will perform the transformation.
Next we derive the T-duals of the NS fields which on their own form a closed set.
This can be done using, for instance, path integral methods following Buscher's treatment of abelian
T-duality \cite{Buscher:1987sk} adapted appropriately for non-abelian isometries \cite{delaossa:1992vc}.
Alternatively, we may achieve the same result by applying a canonical transformation in the phase space
of the two-dimensional $\s$-model \cite{Lozano:1995jx, Sfetsos:1996pm,Curtright:1994be}.
Neither of the above procedures is fully adequate to compute the transformation rules
for the Ramond flux fields.
In  \cite{Sfetsos:2010uq} we developed a general procedure that solved
this problem which is based on the construction
of a Lorentz transformation matrix $\L$ relating the frames naturally
defined by the transformations of the left and right world sheet derivatives under T-duality.

\no
This Lorentz transformation induces an action
on spinors  \cite{Hassan:1999bv}  given by a matrix $\Omega$ obtained by requiring that
\be
\Omega^{-1}  \Gamma^i  \Omega =  \Lambda^i{}_j \Gamma^j\ .
\label{spino1}
\ee
To include RR-fields into the discussion we combine them into
a bi-spinor according to the type--II supergravity to which they belong.
Specifically, we have that
\be
{\rm IIB}:\qq P = {e^{\Phi}\ov 2} \sum_{n=0}^4{1\ov (2n+1)!}\ \slashed{F}_{2n+1} \
\ee
and
\be
({\rm massive)\ IIA}:\qq  P ={ e^{\Phi}\ov 2} \sum_{n=0}^5 {1\ov (2n)!}\ \slashed{F}_{2n} \ ,
\ee
where we used the standard notation
${\slashed F}_p =  \G^{\m_1\cdots \m_p} F_{\m_1\cdots \m_p}$.
In the definition of $P$ we have  used the democratic formulation
of type-II supergravities \cite{Bergshoeff:2001pv} wherein all forms up to order ten appear
on equal footing.
In this formulation and for Minkowski signature spacetimes
the conditions
\be
 F_{2n} = (-1)^n\star F_{10-2n}\ ,\qq
F_{2n+1} = (-1)^n\star F_{9-2n}\ ,
\label{hdhi2}
\ee
should be imposed so that one remains with the right degrees of freedom.
However, in checking our solutions to supergravity we shall, in general,
work with the standard formulations of type-II supergravities in which
no higher forms than five appear.

\no
The Ramond fluxes then transform according to
\be
\hat P = P \Om\inv\ ,
\label{ppom}
\ee
where we have denoted by a hat the bi-spinor obtained after the duality.
In some sense, this relation asserts that, demanding independence of Physics
on the frame choice leads to a tranformation of the flux fields within the two-member
family of type--II supergravity.
The details of the matrix $\Omega$ corresponding to cases of non-abelian T-duality have to be
worked out in the various cases of interest. We recall for comparison that for the case of abelian T-duality
this is simply given as $\Om = \G_{11} \G_1$ \cite{Hassan:1999bv},
where the $1$ labels the isometry direction and $\G_{11}$ the product
of all Gamma matrices. In the abelian case we go from IIA to IIB and vice-versa.
However, in non-abelian cases we might change or stay within the same chirality
theory \cite{Sfetsos:2010uq}.

 \section{Non-abelian T-duals in coset spaces}

In \cite{Sfetsos:2010uq} it was shown that the Lorentz rotation that
acts on spinors can be calculated from the transformation rules of the world sheet derivatives.
These rules are easily obtained in the canonical approach to T-duality.
We now want to understand the same construction
for the coset space $\s$-models.

\subsection{Review of T-duals in group spaces }

We first recap the
results of \cite{Sfetsos:2010uq} which we generalize slightly to incorporate
a wider class of $\s$-models on group manifolds than just the Principal Chiral Model (PCM).
Consider an element $g$ in a group $G$. We construct the components of the left invariant Maurer--Cartan
forms as $L^a_\m =-i\ {\rm Tr}(t^a g^{-1}\del_\m g)$, where the representation matrices
$t^a$ obey the corresponding Lie algebra with structure constants $f^{ab}{}_c$.
The most general $\s$-model, that is invariant under the global symmetry $g\to g_0 g$, with $g_0\in G$, is
of the form
\be
S = \frac{1}{2} \int d^2 \s   \, E_{ab} L^a_+ L^b_-\ ,\qq
L^a_\pm = L^a_\mu \partial_\pm X^\m\ ,
\label{dualmods}
\ee
where $E$ is a $\dim(G)$ square invertible constant matrix
(actually $E$ may depend on other coordinates that have
only a spectator r\^ole in the whole discussion, although this will not
be needed for our purposes). For the case where $E$ is proportional to
just the Cartan metric, taken to be the identity matrix in this paper, this $\s$-model is just the PCM
on $G$. However in what follows it will be important to us that one can still
 perform a duality for a general matrix $E$.

\no
The non-abelian T-dual $\s$-model to \eqref{dualmods} with respect to the full $G$ symmetry
is constructed by following the standard
Buscher-like approach by introducing gauge fields and a Lagrange multiplier term.
Alternatively, we may employ a canonical transformation in phase space.
With either method the result is
\be
\tilde{S} =  \frac{1}{2} \int d^2 \s \,  (M^{-1})^{ab} \partial_+ v_a \partial_- v_b\ ,
\label{dualmodss}
\ee
in which
\be
M_{ab}  = E_{ab} +  f_{ab}\ ,\qq f_{ab}=f_{ab}{}^c v_c\ .
\ee
There is also a dilaton induced as a quantum effect given by
\be
\Phi = -\ha \ln \det M\ .
\label{hfdkl3}
\ee
The canonical transformation relating these models is entirely encoded in the transformation of the
world sheet derivatives
\be
\label{cts}
L_+^a = (M^{-1})^{ba} \del_+ v_b\ ,\qq L_-^a = -(M^{-1})^{ab} \del_- v_b\ .
\ee
As an immediate consequence of the identity
\be
\ha (M^{-1}+ M^{-T}) = M^{-T}\eta M^{-1} = M^{-1}\eta  M^{-T}\ ,
\ee
in which $\eta$ denotes the symmetric part of $E$, both $M^{-1}$ and $M^{-T}$ occurring
in  \eqref{cts} define frame fields for the metric of the dual   $\s$-model
\eqref{dualmodss}. These two frames are related by a Lorentz transformation
\be
\L = -\k M^{-T} M\k^{-1} = - \k^{-T} M M^{-T}\k^T \ ,
\ee
where the matrix $\k$ is such that the constant matrix $\eta= \k^T \k$.
Given this form of the Lorentz transformation we may explicitly solve \eqref{spino1}
to find the corresponding spinorial representation $\Om$.
We first expand $M$ around minus the identity by treating as small parameters the coordinates
$v_a$ as well as the antisymmetric part of the matrix $E$ which we will denote by $S$.
After determining the infinitesimal transformation and subsequent exponentiation we find that
\be
\Om = e^{\ha \tilde f_{ab} \G^{ab}} \prod_{i=1}^{\dim(G)}\! (\G_{11} \G_i) \ ,
\qq \tilde f= \k^{-T} (S+f) \k^{-1} = - \tilde f^T\ .
\label{dhfi}
\ee
The reason that we may obtain the result by an exponentiation of the
infinitesimal form is that the matrices
$\G_{ab}$ close into an $so(\dim(G))$ algebra. From the above expression it is
clear that if the duality group is even
then we stay in the same type-II supergravity, whereas if it is odd then we flip
from (massive) type--IIA
supergravity to type--IIB and vice versa.

\no
Whilst generically the $\s$-model \eqref{dualmodss} has no isometries it
is possible for particular forms of the matrix $E$ to obtain residual symmetries. These correspond
to extra isometries of the original $\s$-model \eqref{dualmods} that commute with the symmetry that
we used to perform the non-abelian T-duality. Of course, the matrix $\Om$ in \eqref{dhfi}
should respect this symmetry. For example, in the case of $E= \mathbb{1}$ the
original $\s$-model in \eqref{dualmods} enjoys a global $G_L\times G_R$
isometry which will lead to a residual $G_R$ symmetry in the dual theory.
This is indeed the case in the examples worked out in \cite{Sfetsos:2010uq}
in which a non-abelian dual of $S^3$ is performed with respect to $SU(2)_L$ of the
total isometry group $SO(4) \simeq  SU(2)_L \times SU(2)_R$; the $SU(2)_R$
symmetry is manifestly preserved in the dual background.

\subsection{Non-abelian T-duals in coset spaces via reduction}

To extend the discussion for $\s$-models corresponding to coset $G/H$ spaces
we split for notational purposes the index $a=(i,\a)$, where the indices $i$ and $\a$ belong
to the subgroup $H\in G$ and the corresponding coset $G/H$, respectively.
The $\s$-model is
\be
S = \frac{1}{2} \int d^2 \s   \, (E_0)_{\a\b} L^\a_\mu L^\b_\nu \partial_+ X^\mu \partial_- X^\nu\ ,
\label{dualmodsco}
\ee
so that it has the same form as that for group spaces in \eqref{dualmods}. The restriction of
the matrix $E$ in \eqref{dualmods} to coset space requires that $E_0$ is $G$-invariant which severely
restricts its form. In most cases of interest this will be taken to be proportional to the Killing metric.
The key point that enables one to obtain the explicit form \eqref{dualmodss} of the non-abelian T-dual
for the case of group manifolds relied on the
fact that the symmetry acts with no isotropy. In technical terms that means that, in the Buscher-like
approach, one can gauge fix the group element $g$ to unity, so that the dual $\s$-model contains only
the Lagrange multipliers. For coset models this is not possible and one has to
gauge fix some of the Lagrange multipliers as well, in which the group acts with isotropy, i.e. as
$\d v^a = f_{bc}{}^a \e^b v^c$. Hence there exist fixed points of this transformation.
For our purposes it is convenient to proceed by using a reduction method
introduced in \cite{Sfetsos:1999zm}.\footnote{This
was actually considered in the more general context of $\s$-models related by Poisson--Lie T-duality,
of which non-abelian duality is just a particular case.}
The reduction procedure is taken as follows: Consider a matrix $E$ of the form
\be
E =\diag\left( E_0,\l\ \mathbb{1}_{\dim{(H)}}\right)\ ,
\ee
where $E_0$ is a $\dim(G/H)$ square invertible constant matrix and $\l$ is a parameter.
Then the dual models \eqref{dualmods} and \eqref{dualmodss} are perfectly consistent and have
$\dim(G)$ target spaces.
In the limit $\l\to 0$ the Maurer--Cartan forms associated with the subgroup
in \eqref{dualmods} drop out. Then, we are left with
the $\s$-model for the coset space $G/H$ \eqref{dualmodsco} and \eqref{dualmodss} represents its dual.
For the whole procedure to be consistent one has to ensure that the corresponding
 target spaces are reduced to
$\dim(G/H)$. It can be shown that this is ensured if $E_0$ is indeed $G$-invariant \cite{Sfetsos:1999zm}.
The above remarks imply that we may fix $\dim(H)$ among the $v_a$'s and
denote the remaining ones by $x_\a$.
Alternatively, we may think of the $x_\alpha$'s as the $H$-subgroup invariants
one can form using the $\dim(G)$
variables parameterizing $g\in G$. This is completely analogous and in fact inspired
by a treatment of the gauge fixing procedure in gauged WZW models in \cite{Bars:1992ti}.

\no
To find out the transformation rules of the world sheet derivatives we define
 the $\dim(G/H)$ square matrices $N_\pm$ from the relations
\begin{eqnarray}
\label{djkh1}
&& L^\a_+ = (M^{-1})^{b\a} \del_+ v_b =  N_+^{\a\b} \del_+ x_\b\ ,
\nonumber
\\
&&
L^\a_- =  -(M^{-1})^{\a b} \del_- v_b = N_-^{\a\b} \del_- x_\b\ ,
\end{eqnarray}
where we have taken the $\l\to 0$ limit.
Then the Lorentz transformation is given by
\be
\L = \k_0 N_+ N_-^{-1}\k_0^{-1} \ ,
\label{lool}
\ee
where $\k_0$ is the restriction of the frame matrix $\k$ to the coset obeying $E_0 = \k_0^T \k_0$.
It should be possible to obtain $\Om$, to be used in \eqref{ppom},
by appropriately taking the $\l\to 0$ limit in \eqref{dhfi}. In that respect, whether or not
one changes or stays in the same type--II theory depends entirely on $\dim(G)$ and not on $\dim(G/H)$.

\section{ Examples}

We present below several examples from D-brane configurations in string theory and
from some standard compactifications in type-II supergravity.

\subsection{Non-abelian T-dual in the D1-D5 near horizon}

As a first example we consider the $AdS_3\times S^3\times T^4$ geometry that
arises as the near horizon limit of the D1-D5 brane system.
The type-IIB supergravity background consists of a metric
\be
ds^2 = ds^2({\rm AdS_3}) + ds^2({\rm S^3}) + ds^2({\rm T^4})\ ,
\ee
where the normalization is such that $R_{\m\n} = \mp 2 g_{\m\n}$ for the $AdS_3$ and $S^3$
factors, respectively, supported by the Ramond flux
\be
F_3 = 2\left({\rm Vol}(AdS_3) + {\rm Vol}(S^3)\right)\ ,
\label{adsfl}
\ee
whereas the dilaton $\Phi =0$. To construct the bi-spinor of fluxes we need
the Hodge-dual of the above three-form
\be
F_7 = -(\star F_3) = 2 \left({\rm Vol}(S^3)+ {\rm Vol}(AdS_3)\right)\wedge {\rm Vol}(T^4) \ .
\ee
Note that we have completely absorbed all constant factors by appropriate rescalings.
The presence of $S^3$ indicates a global $SO(4)$ with respect to which we will perform
the non-abelian transformation.
For comparison, we recall that the non-abelian T-dual with respect
to the $SU(2)_L$ subgroup of $SO(4)$ was constructed in \cite{Sfetsos:2010uq}. However,
in that case, unlike here, the group's action is without isotropy.

\no
To proceed we need to determine the matrix $M$ in \eqref{dualmodss}.
Let's recall that we may construct the $SO(N)$ algebra by first defining matrices
$t_{ab}$ with $a=1,2,\dots , N$, with
\be
(t_{ab})_{cd}= \d_{ac} \d_{bd}\ .
\ee
Then
\be
J_{ab}= t_{ab} - t_{ba}\ ,
\label{jab}\ee
obey the $SO(N)$ algebra. An $SO(N-1)$ subalgebra is generated by the
matrices $J_{ij}$ with $i=2,3,\dots , N$,
whereas the coset $SO(N)/SO(N-1)$ currents are given by $J_{1i}$.

\no
For the case at hand,  $N=4$,  we define
\ba
&& S_a = J_{1,a+1}\ ,\qq a=1,2, 3\ ,
\nonumber\\
&&  S_{a+3} = J_{2,a+2}\ ,\qq a=1,2\ ,
\\
&& S_{6} = J_{34}\ .
\nonumber
\ea
In this arrangement the elements $S_a$ with $a=4,5,6$ obey an $SO(3)$ subalgebra.
We organize the structure constants by computing
\be
[S_a,S_b]=f_{ab}{}^c S_c\ .
\label{reafstr}
\ee

\no
According to the previous discussion we now need to gauge fix three of the six $v_a$.
 For this simple case one could, of course, do this just by inspection. However,
 for more complicated cases this is not such an easy thing to do. To this end we employ
 some group theoretical reasoning developed
in the context of gauged WZW models in \cite{Bars:1992ti}.
Under $SO(4) \rightarrow  SO(3)$ the adjoint decomposes $\bf{6 \rightarrow 3  \oplus
3 }$.  If we label the first triplet as $X$ and the second as $Y$, we have explicitly
\be
X= (v_1 + v_6 ,  v_5-v_2 ,  v_3 +v_4)  \ ,\qq
Y=( v_1 - v_6 ,  -v_5-v_2 ,   v_3  - v_4 )\ .
\ee
There are three independent invariants under $SO(3)$ that one can construct
 from these triplets given by
\be
t_1= X^2\ ,\qq  t_2 = X\cdot Y\ ,\qq t_3 = Y^2\ .
\label{inv3d}
\ee
To fix the residual $SO(3)$ one imposes some constraints $F_i(v)=0$, with $i=1,2,3$. Clearly a
valid gauge fixing choice cannot eliminate these invariants. In other words,
after gauge fixing there must remain three parameters in one-to-one
correspondence with these invariants.
We now make the following gauge choice $v_1=v_2=v_6=0$, and rename the remaining
 coordinates
\be
(x_1,x_2,x_3)= (v_3,v_4,v_5)\ ,
\ee
such that the invariants are given by
\be
t_1 = (x_1 + x_2)^2 + x_3^2\ ,
\quad
t_2= x_1^2 - x_2^2 - x_3^2\ ,
\quad
t_3 = (x_1 - x_2)^2 + x_3^2 \ .
\label{eqttt}
\ee
To construct the dual we now need the matrix $M = E + f$, which
in the $\l \rightarrow 0$ coset limit is given by
\be
M = \left(  \begin{array}{cccccc}
    1 & -v_4  & -v_5  & v_2  & v_3  & 0 \\
    v_4 & 1 &- v_6 & -v_1  & 0& v_3  \\
    v_5  & v_6  & 1 & 0 & -v_1 &-v_2  \\
    -v_2 & v_1  & 0& 0 & -v_6 & v_5  \\
    -v_3 & 0 & v_1  & v_6 & 0 & - v_4  \\
    0 & -v_3 & v_2  & -v_5 & v_4  & 0\\
  \end{array} \right)\ .
\ee
Applying the gauge fixing we find the matrices $N_\pm$ appearing in
the canonical transformation of the derivatives \eqref{djkh1} as
\be
N_+ ={1\ov x_1 x_3} \left(
       \begin{array}{ccc}
         0 & x_2 & x_3 \\
         0 & x_2^2-x_1^2 & x_2 x_3 \\
         x_1 x_3 & x_2 x_3 & x_3^2 \\
       \end{array}
     \right)\ ,
\quad
N_- ={1\ov x_1 x_3} \left(
       \begin{array}{ccc}
         0 & x_2 & x_3 \\
         0 & x_1^2-x_2^2 & -x_2 x_3 \\
         -x_1 x_3 & -x_2 x_3 & -x_3^2 \\
       \end{array}
     \right)\ .
\ee
These define two frames for the dual geometry whose metric is explicitly given by
\ba
\label{mmetr}
ds^2&=&d x_1^2
+2 \frac{ dx_1 ( x_2  dx_2 + x_3  dx_3)}{  x_1}+\frac{ dx_2^2 \left[ x_1^4 - 2 {x_1}^2 x_2^2
+ x_2^2 \left( x_2^2+ x_3^2+1\right)\right]}{  x_1^2 x_3^2} \nonumber \\
&&+ 2 \frac{ dx_2  dx_3  x_2  \left(- x_1^2+ x_2^2
+ x_3^2+1\right)}{ x_1^2 x_3}+\frac{ {dx_3}^2 \left( {x_2}^2+ {x_3}^2+1\right)}{  {x_1}^2}\ ,
\ea
plus of course the terms $ ds^2({\rm AdS_3}) + ds^2({\rm T^4})$.
The NS two-form vanishes and the dilaton computed from \eqref{hfdkl3} is
\be
\Phi =- \ln (x_1 x_3)\ .
\label{siillaaa}
\ee
The Lorentz transformation relating the frames is found using \eqref{lool} with $\k_0=\mathbb{1}$.
It reads
\be
\L =\diag(1,-1,-1)\ .
\ee
Hence the corresponding transformation for the spinors is
\be
\Om = - \G_2 \G_3\ ,
\ee
as if we had two successive abelian T-dualities. The reason for this is
that the lack of isometries in the T-dual background prevents $\Om$ from having some non-trivial
structure.\footnote{In that respect one can check out $\Omega$ in
eq. (3.10) of \cite{Sfetsos:2010uq}. In that case there is a residual rotational symmetry
after the T-duality is performed, so that the matrix $\Om$ could have this symmetry.}
Then we compute the RR forms
\ba
&& F_1 =2  x_1 x_3 e_1 =2( x_2 dx_3 + x_3 dx_3)\ ,
\nonumber\\
&& F_5 = (1+ \star)(F_1 \wedge {\rm Vol}(T^4))\ ,
\label{foo3d}
\ea
supplemented by an $F_9$ obeying $\star F_9 = F_1$ as it should.

\no
The metric \eqref{mmetr} is quite complicated. It turns out that it considerably
simplifies if we use the invariants \eqref{inv3d} as coordinates for the dual geometry.
After some manipulations we find that the natural one-forms
associated with $N_+$ can be expressed quite simply as
\ba
e^1 &=& \frac{1}{8 x_1 x_3} \left(dt_1 - 2 dt_2  +dt_3 \right)\ ,
\nonumber \\
e^2 &=& \frac{1}{4 x_1 x_3} \left[ (x_2- x_1 ) dt_1     + (x_2 + x_1)  dt_3 \right]\ ,
\\
e^3 &= & \frac{1}{4 x_1  } \left( dt_1 + dt_3 \right)\ , \nonumber
\ea where for the time being we leave these $x_\alpha$'s as
implicit functions of the new coordinates (they can be explicitly
obtained by inverting \eqref{eqttt}). For the metric we find
 \ba
 ds^2  & = &  \frac{1/16}{t_1 t_3 - t_2^2 }\
\Big[ (-2 dt_2 + dt_3)^2  + 4 t_1 dt_3^2 + 2dt_1 (-2 dt_2 +  (1-4 t_2) dt_3 )
\nonumber\\
&& \phantom{xxxxxxxx} + (1+4t_3) dt_1^2  \Big] + ds^2(AdS_3) + ds^2(T_4)\ .
 \ea
The dilaton and the fluxes are
\be
  \Phi =  - \ln\left[\ha\sqrt{t_1 t_3 - t_2^2 }\right]\
  \ee
and
\be
F_1 =\frac{1}{4} (dt_1 - 2 dt_2 + dt_3 )\ , \qq F_5= (1+ \ast ) (F_1 \wedge {\rm Vol}(T^4))\ .
\ee

\no
Note that the geometry is singular at $x_1 x_3=0$. This is due to the fact that the
duality group acts with isotropy on the Lagrange multipliers.  In addition, we have verified that the supergravity
equations of motion are indeed satisfied by the T-dual background.
Similar comments apply to the other examples below.

\subsubsection{(No) Supersymmetry of the dual}

The dual background given by \eqref{mmetr}, \eqref{siillaaa} and \eqref{foo3d}
does not preserve any supersymmetry.
This can easily be seen from the dilatino variation (non-democratic form)
 \be
 \delta \l  =
\left( \slashed{\partial}\phi  + i e^\phi \slashed{F}_1 \right)\e
- \frac{1}{2} \left( \slashed{H} + i e^\phi \slashed{F}_3 \right)\e^\ast \ ,
 \ee
 in which $\e = \e_1 + i \e_2$ for two Majorana--Weyl supersymmetry parameters of same chirality.
 For the geometry above, in which we have vanishing three NS form, this simply
 reduces to an equation of the form
 \be
( a \G^{3} + b \G^4 +  c \G^5)  \e = 0\ ,
 \ee
where $a,b,c$ have some coordinate dependence. By squaring one can see
that this implies $(a^2 + b^2 + c^2)\e = 0$ and hence the only solution is the trivial one $\e = 0 $.

\no
This conclusion agrees with our expectation from the spinor-Lorentz-Lie derivative
(Kosmann derivative) \cite{Kosmann,Ortin:2002qb}.
It was shown in \cite{Sfetsos:2010uq} that for the
Killing spinor of $AdS_3 \times S^3 \times T^4$ to be invariant
under the $SU(2)_L$ Killing vectors
\be
({ \cal P}_- \otimes \mathbb{1}_{32} ) \varepsilon = 0\ ,
\ee
where we used the doublet $\varepsilon =  \left( \begin{array}{c}
 \e_1 \\
    \e_2 \\
  \end{array} \right)$ and introduced projectors
$ { \cal P}_\pm = \frac{1}{2} \left( \mathbb{1}_2  \pm \s_1 \right)$.
The fact that the projector for the left action is $\cP_-$ can be traced to  the $SU(2)_L$ invariant 1-forms (and corresponding dual vector fields ) which
 obey the Maurer--Cartan equations
 \be
 dL^a =  \frac{1}{2} f_{bc} {}^a L^b \wedge L^c \, .
 \ee
 If we were instead to consider the $SU(2)_R$ action, since the right invariant forms obey
  \be
 dR^a = -  \frac{1}{2} f_{bc} {}^a R^b \wedge R^c \ ,
 \ee
 we would find a projector condition
    \be
  ({ \cal P}_+ \otimes \mathbb{1}_{32} ) \varepsilon = 0\ .
  \ee
It is clear that the only spinor that can be
invariant under both the left and right actions is the trivial zero spinor.

\subsection{Non-abelian T-dual in the D3 near horizon}

Our second example concerns the type-IIB supergravity solution describing the near horizon limit
of the D3-brane background. It consists of a metric
\be
ds^2 = ds^2({\rm AdS_5}) + ds^2({\rm S^5}) \ ,
\ee
normalized such that $R_{\m\n}= \mp 4 g_{\m\n}$ for the $AdS_5$ and $S^5$ factors, respectively,
supported by the self-dual Ramond flux
\be
F_5 = 4 \left({\rm Vol}(AdS_5) -  {\rm Vol}(S^5)\right)\ .
\ee
As before the dilaton $\Phi = 0$ and we note that we have completely absorbed all constant factors by appropriate rescalings.
The presence of $S^5$ indicates a global $SO(6)$ with respect to which we will perform
the non-abelian transformation.

\no
We construct the $SO(6)$ algebra as in \eqref{jab} with $N=6$ and we define
\ba
&& S_a = J_{1,a+1}\ ,\qq a=1,\dots , 5\ ,
\nonumber\\
&&  S_{a+5} = J_{2,a+2}\ ,\qq a=1,\dots , 4\ ,
\nonumber\\
&& S_{a+9} = J_{3,a+3}\ ,\qq a=1,2, 3\ ,
\label{fhj1}
\\
&& S_{a+12} = J_{4,a+4}\ ,\qq a=1,  2\ ,
\nonumber\\
&& S_{15} = J_{56}\ .
\nonumber
\ea
In this arrangement the elements $S_a$ with $a=6,7,\dots , 15$ obey an $SO(5)$ subalgebra.
We organize the structure constants by computing \eqref{reafstr}.

\no
In order to gauge fix we find it convenient to form the five invariants of the antisymmetric matrix
rep. ${\bf 15}$ of $SO(6)$ under the $SO(5)$ subgroup. According to \eqref{fhj1} this splits into
a vector and the antisymmetric rep., i.e. ${\bf 15\to 5 \oplus 10}$. These are explicitly constructed as
\be
(V_i) =(v_1,v_2,\dots , v_5)\ ,\quad (A_{ij}) = \left(
                                     \begin{array}{ccccc}
                                       0 & v_6 & v_7 & v_8 & v_9 \\
                                       -v_6 & 0 & v_{10} & v_{11} & v_{12} \\
                                       -v_7 & -v_{10} & 0 & v_{13} & v_{14} \\
                                       -v_8 & v_{11} & -v_{13} & 0 & v_{15} \\
                                       -v_{9} & -v_{12} & -v_{14} & -v_{15} & 0 \\
                                     \end{array}\
                                   \right)\ .
\ee
The invariants are\footnote{The characteristic polynomial satisfied by the matrix $A$,
according to the Cayley--Hamilton theorem, is of degree 5. However, because
of antisymmetry of $A$, we have that $(A^5)_{ij}V_i V_j =0$. Hence, the next available invariant
$(A^4)_{ij}V_i V_j $ is not an independent one.
}
\ba
&& t_1= V^2\ ,\qq t_2= -\ha {\rm Tr}(A^2)\ ,\qq t_3 = {1\ov 8}  \e^{ijklm}A_{ij}A_{kl}V_m\ ,
\nonumber\\
&&  t_4 = -{1\ov 4} {\rm Tr}(A^4) +{1\ov 8} [{\rm Tr} (A^2)]^2\ ,\qq t_5 = - (A^2)_{ij} V_i V_j \ ,
\label{finv}
\ea
where the various numerical factors have been introduced for later convenience.
The gauge fixing of ten parameters among the fifteen $v_a$'s  should be such that
the remaining five have a one to one correspondence to the above invariants.
The transformation of the Lagrange multipliers is given by
\be
\d v^a = f_{bc}{}^a \e^b v^c\quad \Longrightarrow \quad \d v^i = f_{jk}{}^i \e^j v^k\ ,\quad \d v^\a = f_{j\b}{}^\a \e^j v^\b\ ,
\ee
where we note that the infinitesimal parameters belong to the subgroup.
One may explicitly check that \eqref{finv} indeed remain invariants.
We choose to keep non-zero the variables
$v_a$ with $ a=5,8,10,13,15$.
Hence in our notation
\be
(x_1,x_2,x_3 ,x_4,x_5)=(v_5,v_8,v_{10},v_{13},v_{15})\ .
\ee
In terms of our
variables $x_\alpha, \alpha=1,2,\dots,5$, the invariants become
\ba
&& t_1 = x_1^2\ ,\qq t_2=x_2^2+x_3^2+x_4^2+x_5^2\ ,\qq t_3 = x_1 x_2 x_3\ ,
\nonumber\\
&& t_4 = x_3^2 ( x_2^2+ x_5^2)\ ,\qq t_5 = x_1^2 x_5^2\ .
\ea
Then the matrix $M$ significantly simplifies.
The matrices that define the frames are computed using \eqref{djkh1}. They
turn out to be
\be
N_+ = {1\ov x_1} \left(
        \begin{array}{ccccc}
          0 & {x_1^2-x_2^2\ov  x_5} & {x_2(x_1^2-x_3^2)\ov  x_3 x_5} & -{x_2 x_4\ov
 x_5} & -x_2 \\
          0 & 0 &   {x_3^2-x_2^2 -x_5^2\ov  x_4 x_5} & {x_3\ov x_5} & 0 \\
          0 & -{x_2 x_4\ov x_5} &
{x_1^2 (x_3^2 - x_2^2) +
 x_3^2 (x_2^2 - x_3^2 - x_4^2 + x_5^2)\ov  x_3 x_4 x_5}
 & {x_1^2-x_3^2-x_4^2\ov  x_5} & -x_4 \\

          0 & -{x_2\ov x_5} & -{x_3\ov  x_5}  & -{x_4\ov  x_5} & -1 \\
          x_1 & x_2 & x_3 & x_4 & x_5 \\
        \end{array}
      \right)\ ,
\ee
as well as a similar expression for $N_-$ in such a way that the
Lorentz transformation \eqref{lool} (we use that $\k_0=\mathbb{1}$) is
\be
\L =\diag(-1,1,-1,1,-1)\ .
\ee
The metric is obtained using either of the above frames. The NS two-form turns out to be zero and the
dilaton is
\be
\Phi = -\ln(x_1^2 x_3 x_4 x_5^2)\ .
\ee
The corresponding transformation for the spinors is (we omit an overall sign)
\be
\Om =  \G_{11} \G_1 \G_3 \G_5\ ,
\ee
leading to the RR form
\be
F_2 = 4 x_1^2 x_3 x_4 x_5^2\ e_2 \wedge e_4\ ,
\ee
together with an $F_8$ obeying $\star F_8 = -F_2$ as it should.

\no
As before we may express the background in terms of the invariants in \eqref{finv}.
After some manipulations we find that the natural one-forms
associated with $N_+$ can be expressed quite simply as
\ba
e^1 &=& \frac{1}{2 x_1^2 x_3 x_5} \left(2 t_1 dt_3 -t_3 dt_1 - t_3 dt_2 \right)\ ,
\nonumber \\
e^2 &=& \frac{1}{2 x_1^3 x_3 x_4 x_5^3} \left[ ( t_1 t_4-t_3^2) dt_2 - t_5 dt_4\right]\ ,
\nonumber\\
e^3 &= & \frac{1}{2 x_1^3 x_3 x_4 x_5^3  }
\Big[ t_4 t_5 dt_1 + (t_2 t_3^2 + t_1^2 t_4 - t_1 (t_3^2 + t_2 t_4) + t_4 t_5) dt_2
\nonumber\\
&&\phantom{xxxx} -
  2 t_3 t_5 dt_3 - t_3^2 dt_4 + t_1 t_4 dt_4 + t_3^2 dt_5 - t_1 t_4 dt_5) \Big]\ ,
\\
e^4 & = & -{1\ov 2 x_1 x_5} dt_2\ ,
\nonumber\\
e^5 & = & {1\ov 2 x_1} (dt_1+dt_2)\ .
\nonumber
\ea
where the $x_\alpha$'s are  implicit functions of the new coordinates.
The dilaton and flux are
\be
\Phi = -\frac{1}{2} \ln \left[(t_3^2 +t_2 t_5 - t_1 t_4)(t_1 t_4-t_3^2 ) - t_4 t_5^2 \right]\ .
\ee
and
\be
F_2 =  - dt_2 \wedge dt_4\ ,
\ee
which is a manifestly exact form.


\no
The involved form of the solution suggests that supersymmetry is broken -- this
is indeed the case as can be established from a consideration of the dilatino
and gravitino supersymmetry variations of type-IIA.  As detailed
in appendix C one reaches the same conclusion by demanding that the spinor-Lie
derivative of the Killing spinors of $AdS_5\times S^5$ vanishes for the $SO(6)$
killing vectors generating the isometry.

 \subsection{\texorpdfstring{Non-abelian T-dual of $AdS_4\times CP_2 \times S^2$ }{Non-abelian T-dual of AdS4CP2S2 }}

There is a class of solutions of eleven-dimensional supergravity labeled as $M(m,n)$,
where $m$ and $n$ are integers,
which were constructed in
\cite{Castellani:1983mf} (for a review see \cite{Duff:1986hr}) and are $U(1)$ bundles
over $CP^2\times S^2$.
By dimensionally reducing one obtains a type-IIA supergravity solution.
The metric is
\be
ds^2 = ds^2({\rm AdS_4}) + ds^2({\rm CP^2}) + ds^2({\rm S^2}) \ ,
\ee
where we have normalized in such a way that
$R_{\m\n}= - 2  g_{\m\n}$, $ \L_4  g_{\m\n} $ and  $\L_2  g_{\m\n} $
 for $AdS_4$, $CP^2$ and $S^2$, respectively.
The geometry is supported by a two-form flux written as a
linear combination of the K\"ahler forms on $CP^2$ and $S^2$
\be
F_2 ={2\ov 3} \L_4\ m J_{CP^2} + \L_2\ n J_{S^2}\ ,
\ee
with
\be
J_{CP^2}= e^1\wedge e^2 + e^3\wedge e^4\ ,\qq J_{S^2}={\rm Vol(S^2)} = e^5 \wedge e^6\ .
\ee
In addition, there is  a four-form flux
\be
F_4 = A_{m,n} {\rm Vol}(AdS_4)\ .
\ee
Consistency with the equations of motion requires that
\be
\L_2= {4\ov 1+2 x}\ ,\qq \L_4= {4 x \ov 1+2 x} \ ,
\ee
and
\be
A_{m,n}^2 = 16 {8 m^2  x^3  - 9 n^2  (1 + x)\ov 9 (x-1) (1+2 x)^2}\ ,
\ee
and that there is a constant dilaton
\be
e^{-2\Phi} ={2\ov 9} {9 n^2 - 4 m^2 x^2\ov (1-x)(1+2 x)}\ .
\label{epco}
\ee
The parameter $x$ is determined from a cubic equation
\be
{m^2\ov n^2} = {9\ov 4} {2 x-1\ov x^2(3 -2 x)}\ ,
\ee
which has only one real root in the interval $x\in [\ha,{3\ov 2}]$. For $x=1$ one easily sees that
consistency requires that $2m=3n$. In this particular case the eleven-dimensional solution has either
$N=2$ or $N=0$ supersymmetry, but the dimensionally reduced type-IIA solution in which we are interested
has no supersymmetry whatsoever, so it is highly unexpected that the T-dual geometry will be supersymmetric.
Note that when $2 m=3 n$ is satisfied then there is
no singularity when $x=1$.

\no
For our purposes we need the higher forms
\ba
&& F_6 = -(\star F_4) = A_{m,n}  \ e^1 \wedge e^2 \wedge e^3 \wedge e^4  \wedge e^5 \wedge e^6 \ ,
\nonumber\\
&& F_8 =\star F_2 = {4\ov 1+2 x}  {\rm Vol (AdS_4)} \wedge \left(
{2\ov 3} m x J_{CP^2}\wedge J_{S^2} +  n  {\rm Vol(CP^2)} \right)\ .
\ea
The presence of $CP^2$ indicates a global $SU(3)$ with respect to which we will perform
the non-abelian transformation. We will use as a basis the standard Gell-Mann
matrices $\l_a$, $a=1,2,\dots, 8$.
To conform with our conventions
we relabel $\l_1, \l_2, \l_3$ and $\l_8$, the generators of the subgroups $SU(2)$ and $U(1)$ respectively,
as $S_5, S_6, S_7, S_8$,
and $\l_4, \l_5, \l_6, \l_7$ as $S_1, S_2, S_3, S_4$.

\no
Now if we consider the symmetry subgroup $SU(2) \times U(1)$, we would like to gauge fix
by setting four of the Lagrange multipliers to zero. A suitable gauge fixing choice may be discerned
 by constructing the four invariants of the $\mathbf{8}$ representation
 of $SU(3)$ under the $SU(2) \times U(1) $
subgroup. Under $SU(2) \times U(1)$, the $\mathbf{8}$
splits as $\mathbf{8} \rightarrow \mathbf{3} \oplus \mathbf{2} \oplus \mathbf{\bar{2}} \oplus \mathbf{1}$.
These may be represented explicitly in terms of the eight Lagrange multipliers as
\be
D = v_8\ , \qq V^i = (v_1 - i v_2, v_3 - i v_4)\ ,\qq  \bar{V}_i = (V^i)^*
\ee
and
\be
A = \left( \begin{array}{cc} v_7 & v_5 - i v_6 \\ v_5 + i v_6 & -v_7 \end{array} \right)\ .
\ee
Then by ensuring that we gauge fix so that the remaining four Lagrange multipliers are in one
to one correspondence with the independent invariants
\be
t_1=D\ ,\qq t_2=\ha {\rm Tr}(A^2)\ , \qq t_3=V^i \bar{V}_i\ ,
\qq t_4 =\ha {\bar V}_i A^i{}_j V^j,
\ee
we will determine a suitable gauge fixing choice. For the current case we adopt $v_2=v_4=v_6=v_7=0$.
This removes any residual freedom in $SU(2) \times U(1)$. We henceforth relabel
\be
(x_1,x_2,x_3,x_4) = (v_1, v_3, v_5, v_8 ).
\ee
For this gauge choice the invariants are
\be
t_1 = x_4\ ,\qq t_2 = x_3^2 \ ,\qq t_3= x_1^2 + x_2^2 \ ,\qq t_4= x_1 x_2 x_3\ .
\ee
The matrices defining the frames may then be read off from the earlier prescription with
$N_+$ taking the form
\be
N_+ = \left(
        \begin{array}{cccc}
          1 & 0 & {x_1(x_2^2-x_3^2) + \sqrt{3} x_2 x_3 x_4\ov x_3 (x_2^2-x_1^2)}
& {x_2(x_2^2+x_1^2-2 x_3^2) + 2 \sqrt{3} x_1 x_3 x_4\ov 2 \sqrt{3} x_3 (x_1^2-x_2^2)} \\
          0 & 0 & {x_2\ov x_2^2 -x_1^2} & {x_1\ov \sqrt{3} (x_1^2-x_2^2)}  \\
          0 & 1 &  {x_2(x_1^2-x_3^2) + \sqrt{3} x_1 x_3 x_4\ov x_3 (x_1^2-x_2^2)}
& {x_1(x_1^2+x_2^2-2 x_3^2) + 2 \sqrt{3} x_2 x_3 x_4\ov 2 \sqrt{3} x_3 (x_2^2-x_1^2)} \\
          0 & 0 & {x_1\ov x_1^2 -x_2^2} & {x_2\ov \sqrt{3} (x_2^2-x_1^2)} \\
        \end{array}
      \right)\ .
\ee
The corresponding Lorentz transformation is
\be
\Lambda = \mbox{diag}(-1,1-1,1) \ ,
\ee
from which one may identify the transformation for the spinors (again omitting an overall sign) as
\be
\Omega = \Gamma_1 \Gamma_3.
\ee
The NS two-form is zero and the dilaton turns out to be
\be
\Phi =  \Phi_0 - \ln \left( 2 \sqrt{3} x_3 (x_1^2-x_2^2) \right)\ ,
\ee
where $\Phi_0$ denotes the constant original dilaton in \eqref{epco}.
The RR fluxes supporting the transformed geometry become
\ba
\label{cp2RR}
F_{2} &=& -{4\ov \sqrt{3}}\ \L_4 m\ x_3 (x_1^2-x_2^2) (e^{2} \wedge e^{3}
+ e^{1} \wedge e^{4})\ ,
\nonumber\\
F_{4} &=&  2 \sqrt{3}\ x_3 (x_1^2-x_2^2)
(\Lambda_2 n \ e^{1} \wedge e^{3} + A_{m,n}\ e^{2} \wedge e^{4} ) \wedge {\rm Vol(S^2)}\ .
\ea
In addition we obtain an $F_6$ and an $F_8$, obeying \eqref{hdhi2}.

\subsection{Non-abelian T-dual of $AdS_4 \times S^2  \times S^2  \times S^2$}

A class of solutions of eleven-dimensional supergravity labeled as $O(n_1,n_2,n_3)$,
where the $n_i$'s are integers, was constructed in
\cite{Nilsson:1984bj} (for a review see \cite{Duff:1986hr}) and are $U(1)$ bundles
over $S^2\times S^2\times S^2$.
By dimensionally reducing one obtains a type-IIA supergravity solution.
The metric is
\be
ds^2 = ds^2({\rm AdS_4}) +  \sum_{i=1}^3 ds^2({\rm S_i^2}) \ ,
\ee
where we have normalized in such a way that
$R_{\m\n}= - 2  g_{\m\n}$ and  $ \L_i  g_{\m\n} $
 for $AdS_4$ and each of the $S^2$'s, respectively.
The geometry is supported by the two-form flux
\be
F_2 = \L_1n_1\ e^1\wedge e^2 +  \L_2 n_2\ e^3\wedge e^4 +  \L_3 n_3\ e^5\wedge e^6\ .
\ee
In addition, there is  a four-form flux which for consistency assumes the form
\be
F_4 = A_{n_1,n_2,n_3}\ {\rm Vol}(AdS_4)\ ,\qq A_{n_1,n_2,n_3}= \sqrt{3}  (\L_1^2 n_1^2 + \L_2^2 n_2^2 + \L_3^2 n_3^2)^{1/2}\ .
\ee
Further, consistency with the equations of motion requires that
\be
\L_1+\L_2+\L_3=4\
\ee
and
\be
{n_1^2\ov n_2^2 }= {\L_2^2 \ov \L_1^2}{ \L_1-1\ov \L_2-1}\ ,\qq {\rm and\ cyclic\ in }\ 1,2,3\  ,
\ee
and that there is a constant dilaton
\be
e^{-\Phi} ={1\ov \sqrt{6}} A_{n_1,n_2,n_3}\ .
\ee
We need the higher forms
\ba
&& F_6 = -(\star F_4) = A_{n_1,n_2,n_3} \ e^1 \wedge e^2 \wedge e^3 \wedge e^4  \wedge e^5 \wedge e^6 \ ,
\nonumber\\
&& F_8 =\star F_2 = {\rm Vol (AdS_4)} \wedge \big(\L_1 n_1\ e^3\wedge e^4 \wedge e^5 \wedge e^6
\\
&& \phantom{xxxx} +  \L_2 n_2\ e^1\wedge e^2 \wedge e^5 \wedge e^6 + \L_3 n_3 \ e^1\wedge e^2 \wedge e^3 \wedge e^4\big)\ .
\nonumber
\ea

\no
We will perform a non-abelian T-duality transformation with respect to the $SU(2)$ symmetry of each one of the
$S^2$ factors.
Let's concentrate on just one of them with metric normalized so that $R_{ij}=g_{ij}$.
We can gauge fix as $v_1 =0$ and
define
\be
(v_2 ,v_3)=(\r,z)\ .
\ee
The matrices defining the frames are
\be
N_+ = \left(
        \begin{array}{cc}
          0 & -{1\ov \r} \\
          1 & {z\ov \r} \\
        \end{array}
      \right)\ ,
\qq
N_- = \left(
        \begin{array}{cc}
          0 & -{1\ov \r} \\
          -1 & -{z\ov \r} \\
        \end{array}
      \right)\ ,
\ee
related by the Lorentz transformation
\be
\L =\diag (1,-1)\ .
\ee
The metric is
\be
ds^2({\rm S_d^2})
 = {d z^2 \ov \r^2}
+ \left(d\r + {z\ov \r}dz\right)^2 \ ,
\ee
whereas the corresponding would be dilaton factor is $\Phi = -\ln \r $ and the
NS two-form is zero.

\no
Taking the above into account we find that the non-abelian dual has metric
\be
ds^2 = ds^2({\rm AdS_4}) + \sum_{i=1}^3 \L_i^{-1} ds^2({\rm S_{d,i}^2})\ ,
\ee
where the $i$-factor contains $(\r_i, z_i)$. The dilaton is
\be
e^{-\Phi} = {1\ov \sqrt{6}} A_{n_1,n_2,n_3}\ \r_1\r_2 \r_3\ .
\ee
To find the non-abelian space requires (we omit again an overall sign)
\be
\Om = \G_{11} \G_2 \G_4 \G_6 \ .
\ee
Hence we obtain
\ba
F_3 & = &  \r_1 \r_2 \r_3 \big(\L_1 n_1\ e^1 \wedge e^4 \wedge e^6
+  \L_2 n_2\ e^2 \wedge e^3 \wedge e^6 +  \L_3 n_3 \  e^2 \wedge e^4 \wedge e^5
\nonumber\\
&& + \ A_{n_1,n_2,n_3}\ e^1 \wedge e^3 \wedge e^5\big)\
\ea
and an $F_7$
obeying $\star F_3 = -F_7$.
Note that from the the original isometry only the permutation symmetry remains and there is no supersymmetry.

\section{Concluding remarks}

In the present paper we have established the rules for performing non-abelian T-duality transformations in cases
where the isometry group acts with isotropy and the supergravity backgrounds
have non-trivial Ramond flux fields. In particular, we have concentrated on
coset spaces that frequently appear in important classical supergravity solutions.

\no
We presented examples starting from D-brane configurations, namely the D1-D5 and the D3 near horizon brane systems,
and also from various supergravity compactifications on spheres and $CP$-spaces. In a similar way to other non-isotropic
cases in \cite{Sfetsos:2010uq} it is possible to stay in the same type-II theory or change chirality from type-IIA
to type-IIB and vice versa,
depending solely on the dimension of the duality group, and irrespectively of the details of the background.

\no
Due to the isotropy there are fixed points of the isometry group acting on the dual variables. These give
rise to singularities in the T-dual backgrounds we have constructed. In addition, as in previous examples,
the T-dual backgrounds correspond to non-compact manifolds  even though the duality groups are compact.
It would be interesting to
investigate possible relations to the near horizon limits of brane configurations.
Then, the singularities could be related to the locations of the branes in the transverse
space.
Another avenue open to investigation is the possibility that our T-dual backgrounds represent
effective theories for describing high spin sectors of some parent theories as it was shown
for pure NS backgrounds in \cite{Polychronakos:2010hd}. If true this will have further implications
within the AdS/CFT correspondence.

\no
Based on our examples, non-abelian T-duality generically breaks all isometries and supersymmetry
when it is performed with respect to the maximal symmetry group.
A further interesting question is to understand whether and how the original
symmetries may be recovered as hidden non-local symmetries in the dual background.

\no
Finally, it would be interesting to derive the same T-duality rules
by dimensional reduction on appropriate manifolds in a similar fashion to the abelian case in \cite{Bergshoeff:1995as}.
For this to be possible one needs to establish relations between compactifications of type-II
supergravity to lower dimensions as well as between their massive deformations.
Besides an alternative proof of the non-abelian T-duality rules in the presence of non-trivial RR fluxes,
this would also provide a deeper understanding of the involved supergravity theories.

\subsection*{Acknowledgements}

\no
We thank K. Siampos for useful discussions. The work of Y.L. has been
supported by the research grants MICINN-09-FPA2009-07122, MEC-DGI-CSD2007-00042 and COF10-03.
D.C.T. thanks CPTH, \'Ecole Polytechnique for its
kind hospitality during a visit where some of this work was completed. He
is supported by the Belgian Federal Science Policy Office through the
Interuniversity Attraction Pole IAP VI/11 and by FWO-Vlaanderen through project G011410N.
E.\'O~C would like to express gratitude not only to IISER Mohali and TIFR, Mumbai, for generous hospitality
during the course of this project, but also to Patta Yogendran for constructive discussion and feedback.


\appendix


\section{ \texorpdfstring{Non-abelian T-dual of $AdS_4\times CP_3$ }{ Non-abelian T-dual of AdS4CP3}}
In this appendix we will examine the type-IIA supergravity solution with metric
\be
ds^2 = ds^2({\rm AdS_4}) + ds^2({\rm CP^3}) \ ,
\ee
normalized such that $R_{\m\n}= - 12 g_{\m\n}$ for the $AdS_4$ and $R_{\m\n}=  8 g_{\m\n} $
for the $CP^3$ factors, respectively. It is
supported by the Ramond fluxes
\be
F_2 =\pm 2 J \ ,\qq F_4 = 6 {\rm Vol}(AdS_4)\ ,
\ee
where $J$ is the K\"ahler form with components obeying $(J^2)_{\m\n} = g_{\m\n}$ (for the $CP^3$ metric
indices only).
The dilaton is $\Phi=0$ and
as before we note that we have completely absorbed all constant factors by appropriate rescalings.

\no
The presence of $CP^3$ indicates a global $SU(4)$ with respect to which we will perform
the non-abelian transformation.
The higher forms are
\ba
&& F_6 = -(\star F_4) = 6\ {\rm Vol(CP^2)} =6\ e^1 \wedge e^2 \wedge e^3 \wedge e^4  \wedge e^5 \wedge e^6 \ ,
\nonumber\\
&& F_8 =\star F_2 = \pm 2 {\rm Vol (AdS_4)} \wedge \big( e^3\wedge e^4 \wedge e^5 \wedge e^6
\\
&& \phantom{xxxx} +   e^1\wedge e^2 \wedge e^5 \wedge e^6 +  e^1\wedge e^2 \wedge e^3 \wedge e^4\big)\ .
\nonumber
\ea
We will denote the generators of the $SU(4)$ algebra by $S_a$, $a=1,2,\dots ,15$ and
we will choose the following anti-hermitian basis \cite{cosetcoinv}
\ba
&& S_1 =\left(
       \begin{array}{cccc}
         0 & 0 & 0 & -i \\
         0 & 0 & 0 & 0 \\
         0 & 0 & 0 & 0 \\
         -i & 0 & 0 & 0 \\
       \end{array}
     \right)\ ,
\qq
S_2 =\left(
       \begin{array}{cccc}
         0 & 0 & 0 & -1 \\
         0 & 0 & 0 & 0 \\
         0 & 0 & 0 & 0 \\
         -1 & 0 & 0 & 0 \\
       \end{array}
     \right)\ ,
\nonumber
\\
&& S_3 =\left(
       \begin{array}{cccc}
         0 & 0 & 0 & 0 \\
         0 & 0 & 0 & -i \\
         0 & 0 & 0 & 0 \\
         0 & -i & 0 & 0 \\
       \end{array}
     \right)\ ,
\qq
S_4 =\left(
       \begin{array}{cccc}
         0 & 0 & 0 & 0 \\
         0 & 0 & 0 & -1 \\
         0 & 0 & 0 & 0 \\
         0 & 1 & 0 & 0 \\
       \end{array}
     \right)\ ,
\label{fhj1cp3a}\\
&&
S_5 =\left(
       \begin{array}{cccc}
         0 & 0 & 0 & 0 \\
         0 & 0 & 0 & 0 \\
         0 & 0 & 0 & -i \\
         0 & 0 & -i & 0 \\
       \end{array}
     \right)\ ,
\qq
S_6 =\left(
       \begin{array}{cccc}
         0 & 0 & 0 & 0 \\
         0 & 0 & 0 & 0 \\
         0 & 0 & 0 & -1 \\
         0 & 0 & 1 & 0 \\
       \end{array}
     \right)\ ,
\nonumber
\ea
and
\be
S_7 = -{i\ov \sqrt{6}}\diag(1,1,1,-3)\ ,\qq S_{7+i} =\left(
                                                       \begin{array}{cc}
                                                         \l_{i} & 0 \\
                                                         0 & 0 \\
                                                       \end{array}
                                                     \right)\ ,\quad i = 1,2,\dots , 8\ ,
\label{fhj1cp3b}
\ee
where $\l_i$ are the Gell-Mann matrices for $SU(3)$.
In this basis the K\"ahler form is \cite{cosetcoinv}
\be
J = e^1 \wedge e^2 + e^3 \wedge e^4+ e^5 \wedge e^6\ .
\ee
Again we organize the structure constants by computing \eqref{reafstr}.
Under the $SU(3)\times U(1)\subset SU(4)$ the $  \bf 15 \rightarrow 8_0 \oplus 3_+ \oplus \bar{3}_- \oplus 1_0 $.
These can be represented in doubled line notation as $A_i{}^j$ , $V^i$,
$\bar{V}_i$ and $D$ where explicitly in terms of the 15 $v_a$'s we have
\be
D= v_7\ , \qq V^i = ( v_1- i v_2 , v_3 - i v_4, v_5 - i v_6) \ ,\qq
\bar{V}_i = (V^i)^\ast
\ee
and
\be
(A^i{}_j) =  \left(
\begin{array}{ccc}
 v_{10}+\frac{v_{15}}{\sqrt{3}} & v_{8}-i v_{9} & v_{11}-i
   v_{12} \\
 v_{8}+i v_{9} & \frac{v_{15}}{\sqrt{3}}-v_{10} & v_{13}-i
   v_{14} \\
 v_{11}+i v_{12} & v_{13}+i v_{14} & -\frac{2 v
   _{15}}{\sqrt{3}}
\end{array}
\right) \,  .
\ee
There are two classes of charge invariant operators that can be built
by forming contractions; "glueballs" of the form ${\rm Tr}(A^n) $ and "mesons"
of the form $ V^i (A^n)_i{}^j  \bar{V_j}$.  However, trace relations similar to
those mentioned in the main text,
ensure that these are not all independent
and a suitable basis is given by
\ba
&&  t_{1}= V^i\bar{V_i}\ ,  \qq t_{2}= \bar V_i A^i{}_j
V^j\ ,   \qq t_{3}=\bar V_i (A^2)^i{}_j  V^j - \frac{1}{2} V^i\bar{V_i}  {\rm Tr}(A^2)
 \nonumber \\
 && t_{4}= \frac{1}{2}{\rm Tr}(A^2)\ ,\qq t_{5}=\frac{1}{3} {\rm Tr}(A^3), \qq t_{6 }  =2 \sqrt{\frac{2}{3}} D \ . \qq
\ea
As already pointed out in the main text,
the gauge fixing of nine parameters among the fifteen should be in one to one correspondence with the
above invariants.
We choose to keep non-zero the variables
$v_a$ with $ a=1,6,7,8,10,12$.
Adopting the notation
\be
(x_1,x_2,x_3 ,x_4,x_5,x_6)=(v_1 , v_6 , v_7 , v_8 , v_{10}, v_{12})\ ,
\ee
the invariants are
\ba
t_{1} &=& x_{1}^{2} + x_{2}^{2}  \ , \qq
t_{2} = x_{1} \left( x_{1} x_{5} -2 x_{2} x_{6}\right) \ ,
\nonumber \\
t_{3} &=&  -x_{2}\left( x_{2} (x_{4}^{2} + x_{5}^{2}) + 2 x_{1} x_{5}x_{6}  \right) \ ,
 \\
t_{4} &=&  x_{4}^{2} + x_{5}^{2} +x_{6}^{2}     \ , \qq  t_{5} =   x_{5}x_{6}^{2}  \ , \qq t_{6}= 2 \sqrt{\frac{2}{3}} x_{3} \ .
 \nonumber
\ea

\no
One could now compute the matrices which define frames by following
 the procedure prescribed in \eqref{djkh1}, however this entails
the onerous task of inverting a large matrix.
An alternative approach described in \cite{Sfetsos:1999zm,Balazs:1997be},
is to start directly with the generating functional of the canonical
 transformation between dual $\s$-models, apply the above gauge fixing and then
calculate the remaining transformations.   Finally one sets to zero the components of
momenta in the direction of the subgroup since these  drop out  of the $\s$-model
 \eqref{dualmodsco} in the coset limit.
In this way one may calculate the following explicit, albeit extremely complicated,
expressions for the frames of the dual $\s$-model
\ba
2 \Delta e^{1} &=&  x_1 s_1 dt_1 + x_2^2 x_6 dt_2 + x_1 x_2 dt_3+  (x_1 s_1 + t_6 x_2^2 x_6)  dt_4 + x_2 s_3 dt_5
\nonumber  \\
&& +\ \frac{1}{4}\left(s_2 s_3 - x_1^3 x_2x_5 - 2 x_2^4 x_6   \right) dt_6 \ ,
 \nonumber \\
2 \Delta e^{2} &=& x_2^2 x_6 dt_4 + x_1 x_2 dt_5 + \frac{1}{4}x_1 s_2 dt_6 \ ,
\nonumber \\
2 x_4 \Delta e^{3} &=&- x_1^{-1} \left(s_5 x_2 x_5 x_6 + s_6 s_7\right)dt_1 + s_6 dt_2 - s_5 dt_3 +\left( s_6 (t_6 -x_5) + x_4^2 x_2^2 x_6 \right)dt_4
 \nonumber  \\
&&  +\ (s_6 - s_5 t_6 )dt_5 +\frac{1}{4}\left(\left(s_4 x_6 - s_7 s_2     \right)  t_6 + s_8\right) dt_6 \ ,
\nonumber \\
2 x_4 \Delta e^{4} &=& s_6 dt_4 - s_5 dt_5 + \frac{1}{4}\left(s_5x_6^2 -2s_6x_5 + (x_1^2+2 x_2^2) x_4^2x_6  \right) dt_6 \ ,
 \\
2  \Delta e^{5}  &=& x_1 x_2 x_6 dt_4 + x_1^2 dt_5 - \frac{s_4}{4}  dt_6 \ ,
\nonumber \\
2   \Delta e^{6}  &=& -x_2^2 x_6^2 dt_1 - x_1 x_2 x_6 dt_2 -x_1^2 dt_3  - x_2 x_6s_3 dt_4 -x_1 s_3 dt_5
\nonumber\\
&&  +\ \frac{1}{4}\left(t_6s_4 + x_1^4x_5 +2 x_1x_2^3 x_6 - x_6 s_2 \right) dt_6\ ,
\nonumber
\ea
in which we have defined
\ba
&& s_1 = t_4 x_2 + x_1 x_5 x_6 \ , \qq s_2 =  2 t_4 x_2 +3  x_1 x_5 x_6 \ ,  \qquad s_3 = t_6 x_1 + x_2 x_6\ ,
\nonumber \\
 && s_4 = t_4 x_1^2 - 3 x_2^2 x_6^2  \ , \qquad  s_5 = x_1 x_2 x_5 + (x_1^2 -x_2^2)x_6 \ , \qquad s_6 = x_5 s_5 + x_1 x_2 x_4^2  \ ,
\nonumber\\
  && s_7 = x_1 x_5 - x_2 x_6  \ , \qquad \Delta = x_1^2 s_1 - x_2^3 x_6^2  \ ,
\nonumber \\
  &&  s_8 = 2 t_4  (s_6 - t_1 x_1 x_2)  + x_1^3 x_2 x_5^2 - (x_1^4 +3 x_1^2 x_2^2 -2x_2^4)x_5x_6 + ( 4x_1x_2^3 - 3 s_5 x_5)x_6^2   \ .
\nonumber
\ea
The dual background has a dilaton given by
\be
\Phi = -\ln(  4\sqrt{2} x_{4} \Delta )\
\ee
and zero NS two-form field.\footnote{Other gauge fixing choices may  result in a
non-zero NS two-form.  However, these will be pure gauge with vanishing field strength.}
Whilst this background as presented is clearly very complicated one might hope that
 some more sophisticated group theoretic arguments could be brought to bear in
  order that it can be understood better.

\no
The Lorentz transformation relating left and right movers is given by
\be
\L =\diag(-1,1,-1,1,1,-1)\ ,
\ee
which has the spinorial representation
\be
\Om = \G_{11} \G_{1}\G_{3}\G_{6}\ .
\ee
Therefore we conclude that the dual geometry is supported by the following flux
\be
F_3 =   \pm 8 \sqrt{2} x_4 \D (3 e^2 \wedge e^4 \wedge e^5 +  e^1 \wedge e^3 \wedge e^5
-e^2 \wedge e^3 \wedge e^6 - e^1 \wedge e^4 \wedge e^6)\ .
\ee
There is also an $F_7 = -(\star F_3)$ as it should.

\section{Geometry and Killing vectors in group and coset spaces}

For the reader's convenience we recapitulate some relevant results for our purposes
concerning the geometry of groups and
coset manifolds.
Further details may be found in \cite{Ferrara:1984ij, Castellani:1983tb}.
Following our notation in the main text,
let $t_a$ be generators for $G$ of which $t_i$ correspond to the subgroup
$H\subset G$ and $t_\a$ are the remaining coset generators.
We assume that the generators are normalised such that $\textrm{Tr}(t_a t_b)= \delta_{ab}$.
An element $g\in G$, parameterized appropriately by $\dim(G)$ variables $X^\m $,
can be used to define the $G$-algebra valued left-invariant and right-invariant
one-forms $L =-i g^{-1} dg = L^a t_a$ and $R =-i g^{-1} dg g^{-1} = R^a t_a$, with components
related by $R^a = D^{ab} L^b$, where
\be
D_{ab}(g)  = {\rm Tr}( g^{-1} t_a g t_b) \ .
\ee
This matrix is defined by the adjoint action of $g$ and obeys $D_{ab}(g^{-1}) = D_{ba}(g)$.
The metric in group space is
\be
g_{\m\n} = L^a_\m L^a_\n = R^a_\m R^a_\n\ .
\ee
This metric has a $G_L\times G_R$ group of invariance.
The Killing vectors for these left and right transformations are
\be
K^{\rm L}_a =  R^\m_a \del_\m \ , \qq K^{\rm R}_a = - L^\m_a \del_\m\ .
\label{B3}
\ee
They obey two commuting Lie-algebras for $G$ as well as a completeness and a derivative relation
\be
\sum_{a= 1}^{\dim (G)}K_a^\m K_a^\n = g^{\m\n} \ , \quad  \nabla_{\m} K^a_\n
= -\frac{1}{2} f_{bc}{}^a K^b_\m K^c_\n \ ,
\label{akfgrs}
\ee
for either set of Killing vectors, separately.
Also $\partial_\m D_{ab} = L_\m^c D_{ad} f_{cb}{}^d$ proves useful in various algebraic manipulations.

\no
Turning to coset spaces, an element of the coset $G/H$ is given by a representative
$\hat{g}\in G$ parameterized by $\dim(G/H)$
local coordinates $x^\m$, for example $\hat{g} = \exp(i t_\a \delta_{\m}^\a x^\m)$.
The left-invariant one-forms with coset indices $L^\a$
define a frame for the coset $e_\m{}^\a = L^\a_\m $, with inverse $e^\m{}_\a$,
such that the metric on the coset is given by
\be
g_{\m \n}= e_\m{}^\a e_\n{}^\b \d_{\a\b} \
\ee
and has only a $G_L$ invariance group.
The corresponding Killing vectors are
\be
K_a =D_{a \a}(\hat{g})e_\a^\m \partial_\m \ ,
\ee
which obey the relations \eqref{akfgrs} with $\m$ and $\n$ parameterizing the coset manifold.
It turns out that $K_a$ can be obtained from the Killing
vectors $K^L_a$ defined in \eqref{B3} for group spaces as follows: Parameterizing
a general group element in $G$ as $g=\hat g h$, where $h\in H$, and denoting the variables parameterizing
$h$ by $y^i$, arranged such that $y^i=0$ corresponds to $h=\mathbb{I}$,
$K_a$ can be obtained from $K^L_a$ by "gauge fixing", i.e. by setting $y^i=0$
and ignoring the corresponding derivatives $\del_i$.

\section{Spinor derivative on $AdS_5\times S^5$ Killing vectors }

The Killing spinors of the $AdS_5 \times S^5$ geometry in type-IIB supergravity obey the differential equation
\be
0 = (\mathbb{1}_2 \otimes D_\m) \e + \frac{1}{8 \cdot 5!} (i \s_2\otimes \slashed{F}_5\G_\m) \e\ ,
\ee
or in terms of the complex Weyl spinor $\varepsilon = \e_1 + i \e_2$,
\be
\label{gravitinoeq}
0= D_\m \varepsilon - \frac{i}{8 \cdot 5!} \slashed{F}_5 \G_\m  \varepsilon  \ .
\ee
One may choose a basis of Gamma-matrices (see the appendix of \cite{Lu:1998nu})
\be
\G_a = \s_1 \otimes \g_a \otimes 1 \ , \qq \G_i = \s_2 \otimes 1 \otimes \g_i \ ,
\ee
for the AdS and sphere directions respectively, such that a chiral spinor may be decomposed as
\be
\varepsilon = \left(\begin{array}{c} 1 \\ 0  \end{array}\right)\otimes \e_{AdS} \otimes \eta \, .
\ee
The components of \eqref{gravitinoeq} in the directions of the sphere (with unit radius) become
\be
D_\a \eta = \frac{i}{2} \g_\a \eta \ , \qq \a=5\dots 9\ ,
\ee
(in which $\eta$ has four complex but otherwise unconstrained components) whilst in the direction of AdS
\be
\label{eq:adsks}
D_a \e_{AdS} = \frac{1}{2} \g_a \e_{AdS}  , \qq a=0,1,\dots , 4\ .
\ee
Then for the Killing vectors generating the $SO(6)$ isometry we may act with the spinor-Lorentz-Lie derivative
as
\ba
{\cal L }_{K^a} \eta &=&   K^{a\a}D_\a \eta - \frac{1}{4}\nabla_\a K^a_\b \g^{\a\b} \eta
\nonumber \\
& =& \frac{i}{2} K^a_\a \g^\a \eta  - \frac{1}{4}\nabla_\a K^a_\b \g^{\a\b} \eta \\
&=& \frac{i}{2} K^a_\a \g^\a \eta  - \frac{1}{4}f_{bc}{}^a  K^b_\a K^c_\b \g^{\a\b} \eta\ .
\nonumber
\ea
In the second line we have used the Killing spinor property and in the third the property obeyed by the
Killing vectors in \eqref{akfgrs}. Next we contract the above equation with $ K^a_\a $
and make use of the completeness relation on the Killing
 vectors in \eqref{akfgrs} to obtain that
\be
\sum_{a=1}^{\dim(G)} K^a_\a {\cal L }_{K^a}  \eta = \frac{i}{2}
\g_\a \eta - \frac{1}{4} f_{abc}K^a_\a K^b_\b K^c_\g \g^{\b\g} \eta\ .
\ee
However, as shown in \eqref{chk1} below, it turns out that
$f_{abc}K^a_\a K^b_\b K^c_\g  = 0 $, $\forall\ \a,\b$ and $\g$.
Hence for a Killing spinor to be invariant under the $SO(6)$  action it is necessary that
\be
\sum_{a=1}^{\dim(G)} K^a_\a {\cal L }_{K^a}  \eta = \frac{i}{2} \g_\a \eta = 0\ ,
\ee
to which the only solution is $\eta=0$. Hence, we conclude that the non-abelian T-dual
of the $AdS_5\times S^5$ background of type-IIB does not preserve any supersymmetry.

\section{$S^5$ as a coset and its $SO(6)$ Killing vectors }

The five-sphere of half-unit radius  can be defined by five stereographic coordinates
$z^\a = y^\a ( 1/2- y^6)^{-1}$ where $\vec{y}$ defines the embedding in $\mathbb{R}^6$.
The isomorphism between the sphere and the coset $SO(6)/SO(5)$ is given by identifying with a
point $\vec{z}$ an $SO(6)$ element that maps the north pole to that point, modulo the $SO(5)$ stability group
which leaves the north pole fixed.  Corresponding to the point $\vec{z}$
one may take as the group element (see, for instance,
 the contribution of Van Nieuwenhuizen in \cite{Ferrara:1984ij})
\be
\hat{g} (\vec{z} )=  (1+ z^2)^{-1}  \left(
       \begin{array}{c|c}
        \d^{\a\b}(1+ z^2) -2z^\a z^\b &  2z^\b   \\
  \hline      - 2z^\a   & 1- z^2 \\
       \end{array}
     \right) \ ,
\ee
in a basis in which the subgroup generators
act on the top left block and the coset acts on the
remaining directions.  With  $(J_{a, b })_{cd} = \d_{a c}\d_{b d} - \d_{a d}\d_{b c}$
the coset generators are then given by $t_{\a}= J_{6 , \a}$, $\a=1,2,\dots, 5$
and the subgroup generators are all the rest.
Following the steps described above one recovers the metric
\be
ds^2 = \frac{4 dz^\a dz^\a }{(1+z^2)^2}\ ,\qq z^2 = z^\a z^\a\
\ee
and finds the Killing vectors to be
\ba
K_{a}&=&  z^{a+1} \partial_1 - z^1 \partial_{a+1 }   \ , \quad a = 1,2 ,3,4\ ,
 \nonumber \\
 K_{a+4} &=& z^{a+2} \partial_2 - z^2 \partial_{a +2}     \ , \quad a = 1,2, 3  \ ,
\nonumber\\
K_{a + 7} &=&  z^{a+3} \partial_3 - z^3 \partial_{a +3}\ , \quad a= 1, 2\ ,
 \\
K_{10} &=& z^5\partial_4 - z^4 \partial_5\ ,
\nonumber\\
K_{a+10}  &=&  z^a z\cdot \partial   + \frac{1-z^2}{2}\partial_a \ , \quad a=1,2,\dots, 5 \ .
\nonumber
\ea
A tedious direct calculation verifies that $f_{abc} K^a_\a K^b_\b K^c_\g = 0 $, $\forall \a,\b$ and $\g$,
where the summation acts
on the entire set of $SO(6)$ algebra indices. Equivalently the three-form
\be
f_{abc} K^a_\a K^b_\b K^c_\g dz^\a  \wedge  dz^\b  \wedge dz^\g = 0\ .
\label{chk1}
\ee
It would be interesting to know to what extent this vanishing relation is valid for other cosets as well.




\begin{thebibliography}{99}

\bibitem{Buscher:1987sk}
T.H. Buscher, {\it {A Symmetry of the String Background Field Equations}},
   Phys. Lett. {\bf B194} (1987) 59 and {\it {Path Integral Derivation of Quantum Duality in Nonlinear
  Sigma Models}},  Phys. Lett. {\bf B201} (1988) 466.

\bibitem{delaossa:1992vc}
X.C. de~la Ossa and F.~Quevedo, {\it Duality symmetries from non abelian
  isometries in string theory},  Nucl. Phys. {\bf B403} (1993) 377,
  \href{http://xxx.lanl.gov/abs/hep-th/9210021}{{\tt arXiv:hep-th/9210021}}.

\bibitem{Giveon:1993ai}
A.~Giveon and M.~Rocek, {\it {On nonAbelian duality}},  Nucl. Phys. {\bf
  B421} (1994) 173, \href{http://xxx.lanl.gov/abs/hep-th/9308154}{{\tt
  arXiv:hep-th/9308154}}.

\bibitem{Alvarez:1994zr}
E.~Alvarez, L.~Alvarez-Gaume and Y.~Lozano, {\it {On nonAbelian duality}},
  Nucl. Phys. {\bf B424} (1994) 155,
  \href{http://xxx.lanl.gov/abs/hep-th/9403155}{{\tt arXiv:hep-th/9403155}}.
\bibitem{Sfetsos:1994vz}
  K.~Sfetsos,
 {\it Gauged WZW models and nonAbelian duality},
  Phys. Rev.  {\bf D50} (1994) 2784,
   \href{http://xxx.lanl.gov/abs/hep-th/9402031}{{\tt arXiv:hep-th/9402031}}.

\bibitem{Lozano:1995jx}
Y.~Lozano, {\it {NonAbelian duality and canonical transformations}},  Phys. Lett. {\bf B355} (1995) 165,
  \href{http://xxx.lanl.gov/abs/hep-th/9503045}{{\tt arXiv:hep-th/9503045}}.

\bibitem{Sfetsos:1996pm}
K.~Sfetsos, {\it {Non--Abelian Duality, Parafermions and Supersymmetry}},
Phys. Rev. {\bf D54} (1996) 1682,
  \href{http://xxx.lanl.gov/abs/hep-th/9602179}{{\tt arXiv:hep-th/9602179}}.



\bibitem{Polychronakos:2010fg}
  A.P.~Polychronakos, K.~Sfetsos,
  {\it Solving field equations in non-isometric coset CFT backgrounds}
  Nucl. Phys. {\bf B840 } (2010)  534,
  \href{http://xxx.lanl.gov/abs/1006.2386}{\tt arXiv:1006.2386 [hep-th]}.

\bibitem{Polychronakos:2010hd}
  A.P.~Polychronakos, K.~Sfetsos,
  {\it High spin limits and non-abelian T-duality},
  Nucl. Phys.  {\bf B843 } (2011)  344,
\href{http://xxx.lanl.gov/abs/1008.3909}{\tt arXiv:1008.3909 [hep-th]}.


\bibitem{Berkovits:2008ic}
N.~Berkovits and J.~Maldacena, {\it Fermionic t-duality, dual superconformal
  symmetry, and the amplitude/wilson loop connection},  JHEP {\bf 09}
  (2008) 062, \href{http://xxx.lanl.gov/abs/0807.3196}{{\tt
  arXiv:0807.3196 [hep-th]}}.

\bibitem{Sfetsos:2010uq}
  K.~Sfetsos and D.C.~Thompson,
  {\it On non-abelian T-dual geometries with Ramond fluxes},
Nucl. Phys.  {\bf B846 } (2011)  21,
\href{http://xxx.lanl.gov/abs/1012.1320}{{\tt arXiv:1012.1320 [hep-th]}}.


\bibitem{Gaiotto:2009gz}
  D.~Gaiotto, J.~Maldacena,
  {\it The Gravity duals of N=2 superconformal field theories},
\href{http://xxx.lanl.gov/abs/0904.4466}{\tt arXiv:0904.4466 [hep-th]}.

\bibitem{ReidEdwards:2010qs}
  R.A. Reid-Edwards and B.J. Stefanski,
  {\it {On Type IIA geometries dual to N = 2 SCFTs} }
 \href{http://xxx.lanl.gov/abs/1011.0216}{{\tt  arXiv:1011.0216 [hep-th]}}.

\bibitem{Bergshoeff:1995as}
E.~Bergshoeff, C.~M. Hull, and T.~Ortin,
{\it {Duality in the type II superstring effective action}},
Nucl. Phys. {\bf B451} (1995) 547, {\tt arXiv:hep-th/9504081}.


\bibitem{Curtright:1994be}
T.~Curtright and C.K. Zachos, {\it {Currents, charges, and canonical structure
  of pseudodual chiral models}},  Phys. Rev. {\bf D49} (1994) 5408,
  \href{http://xxx.lanl.gov/abs/hep-th/9401006}
  {{\tt arXiv:hep-th/9401006}}.

\bibitem{Hassan:1999bv}
S.~F. Hassan, {\it {T-duality, space-time spinors and R-R fields in curved
  backgrounds}},  Nucl. Phys. {\bf B568} (2000) 145,
  \href{http://xxx.lanl.gov/abs/hep-th/9907152}{{\tt arXiv:hep-th/9907152}}.

\bibitem{Bergshoeff:2001pv}
E.~Bergshoeff, R.~Kallosh, T.~Ortin, D.~Roest and A.~Van~Proeyen, {\it {New
  Formulations of D=10 Supersymmetry and D8-O8 Domain Walls}},  Class.
  Quant. Grav. {\bf 18} (2001) 3359,
  \href{http://xxx.lanl.gov/abs/hep-th/0103233}{{\tt arXiv:hep-th/0103233}}.

\bibitem{Sfetsos:1999zm}
  K.~Sfetsos,
  {\it Duality invariant class of two-dimensional field theories},
  Nucl. Phys.  {\bf B561 } (1999)  316,
   \href{http://xxx.lanl.gov/abs/hep-th/9904188}{{\tt arXiv:hep-th/9904188}}.

\bibitem{Bars:1992ti}
  I.~Bars and K.~Sfetsos,
  {\it Global analysis of new gravitational singularities in string and particle theories},
  Phys. Rev.  {\bf D46 } (1992)  4495,
  \href{http://xxx.lanl.gov/abs/hep-th/9205037}{{\tt arXiv:hep-th/9205037}}.

\bibitem{Kosmann}
Y.~Kosmann, {\it {A note on Lie-Lorentz derivatives}},   Annali di Mat.
  Pura Appl. {\bf (IV) 91} (1972) 317.

\bibitem{Ortin:2002qb}
T.~Ortin, {\it {A note on Lie-Lorentz derivatives}},  Class. Quant. Grav.
  {\bf 19} (2002) L143,\hfill\break
  \href{http://xxx.lanl.gov/abs/hep-th/0206159}{{\tt arXiv:hep-th/0206159}}.

\bibitem{Castellani:1983mf}
  L.~Castellani, R.~D'Auria and P.~Fre,
  {\it $SU(3) \times SU(2) \times  U(1)$ from $D = 11$ supergravity},
  Nucl. Phys.  {\bf B239} (1984) 610.

\bibitem{Duff:1986hr}
  M.J.~Duff, B.E.W.~Nilsson and C.N.~Pope,
  {\it Kaluza--Klein Supergravity},
  Phys. Rept.  {\bf 130} (1986) 1.

\bibitem{Nilsson:1984bj}
  B.E.~W.~Nilsson and C.N.~Pope,
  {\it Hopf Fibration Of Eleven-dimensional Supergravity},
  Class. Quant. Grav.  {\bf 1 } (1984)  499.


\bibitem{cosetcoinv}
  F.~Mueller-Hoissen and R.~Stuckl,
{\it Coset spaces and ten-dimensional unified theories},
  Class. Quant. Grav.  {\bf 5} (1988) 27.

\bibitem{Balazs:1997be}
  L.~K.~Balazs, J.~Balog, P.~Forgacs,  N. Mohammedi, L.Palla and  J. Schnittger,
 {\it Quantum equivalence of sigma models related by nonAbelian duality transformations},
  Phys. Rev. {\bf D57} (1998) 3585,
  \href{http://xxx.lanl.gov/abs/hep-th/9704137}{{\tt arXiv:hep-th/9704137}}.


\bibitem{Lu:1998nu}
  H.~Lu, C.~N.~Pope, J.~Rahmfeld,
  {\it A Construction of Killing spinors on S**n},
  J.\ Math.\ Phys.\  {\bf 40 } (1999)  4518-4526,
 \href{http://xxx.lanl.gov/abs/hep-th/9805151}{{\tt arXiv:hep-th/9805151}}.

\bibitem{Ferrara:1984ij}
  S.~Ferrara, (Ed.), J.~G.~Taylor, (Ed.), P.~Van Nieuwenhuizen, (Ed.),
  {\it Supersymmetry And Supergravity '82. Proceedings Trieste School, Italy, September 6-15, 1982},
  Singapore, Singapore: World Scientific ( 1983) 334p.

\bibitem{Castellani:1983tb}
  L.~Castellani, L.~J.~Romans, N.~P.~Warner,
  {\it  Symmetries Of Coset Spaces And Kaluza-klein Supergravity},
  Annals Phys.\  {\bf 157 } (1984)  394.


\end{thebibliography}

\providecommand{\href}[2]{#2}\begingroup\raggedright\endgroup

\end{document}